\documentclass[fleqn,usenatbib]{mnras}

\usepackage{newtxtext,newtxmath}

\usepackage[T1]{fontenc}

\DeclareRobustCommand{\VAN}[3]{#2}
\let\VANthebibliography\thebibliography
\def\thebibliography{\DeclareRobustCommand{\VAN}[3]{##3}\VANthebibliography}

\usepackage{graphicx}	
\usepackage{amsmath}	

\title[Distinguishing a planetary transit signal]{
Distinguishing a planetary transit from false positives: a Transformer-based classification for planetary transit signals
}

\author[Helem Salinas]{
Helem Salinas,$^{1}$\thanks{E-mail: yhsalinas@uc.cl}
Karim Pichara,$^{1}$\thanks{E-mail: kpb@ing.puc.cl}
Rafael Brahm,$^{2,3}$\thanks{E-mail: rafael.brahm@uai.cl}
Francisco P\'erez-Galarce,$^{1}$\thanks{E-mail: fjperez10@uc.cl} and
Domingo Mery,$^{1}$\thanks{E-mail: domingo.mery@uc.cl} \\
$^{1}$Departamento de Ciencias Computacionales, Facultad de Ingenier\'ia, Pontificia Universidad Cat\'olica de Chile, Santiago, Chile\\
$^{2}$Facultad de Ingenier\'ia y Ciencias, Universidad Adolfo Ib\'añez, Av. Diagonal las Torres 2640, Peñalolén, Santiago, Chile\\
$^{3}$Millennium Institute for Astrophysics, Chile
}

\date{Accepted 2023 April 12. Received 2023 March 24; in original form 2022 November 30}
\pubyear{2015}

\begin{document}
\label{firstpage}
\pagerange{\pageref{firstpage}--\pageref{lastpage}}
\maketitle

\begin{abstract}

Current space-based missions, such as the Transiting Exoplanet Survey Satellite (TESS), provide a large database of light curves that must be analysed efficiently and systematically. In recent years, deep learning (DL) methods, particularly convolutional neural networks (CNN), have been used to classify transit signals of candidate exoplanets automatically. However, CNNs have some drawbacks; for example, they require many layers to capture dependencies on sequential data, such as light curves, making the network so large that it eventually becomes impractical. The self-attention mechanism is a DL technique that attempts to mimic the action of selectively focusing on some relevant things while ignoring others. Models, such as the Transformer architecture, were recently proposed for sequential data with successful results. Based on these successful models, we present a new architecture for the automatic classification of transit signals. Our proposed architecture is designed to capture the most significant features of a transit signal and stellar parameters through the self-attention mechanism. In addition to model prediction, we take advantage of attention map inspection, obtaining a more interpretable DL approach. Thus, we can identify the relevance of each element to differentiate a transit signal from false positives, simplifying the manual examination of candidates. We show that our architecture achieves competitive results concerning the CNNs applied for recognizing exoplanetary transit signals in data from the TESS telescope. Based on these results, we demonstrate that applying this state-of-the-art DL model to light curves can be a powerful technique for transit signal detection while offering a level of interpretability.

\end{abstract}




\section{Introduction}


For years, over a million stars have been observed to detect extra-solar planets through a variety of techniques. One of the most successful methods to detect exoplanets is the transit method, which consists of monitoring the flux of stars in the search for periodic dips in brightness. This method considers that those dips could be produced by planets blocking a portion of the host star, as seen by the observer. The CoRoT (Convection, Rotation and planetary Transits) mission \citep{auvergne2009corot}, was the first space-based telescope to detect exoplanets using the transit method. Perhaps the most notable contributor is NASA’s Kepler space telescope \citep{borucki2010kepler}, which was launched in 2009. Kepler identified nearly 4,000 potential planet candidates, out of which $\sim$2,600 have been confirmed as exoplanets to date. Because Kepler could no longer observe its original field, an extended version of Kepler, K2, was designed \citep{howell2014k2}. K2 identified $\sim$500 new planets. From the full sample of 5,200 discovered planets\footnote{https://exoplanetarchive.ipac.caltech.edu/index.html}, most have been identified as candidates with the transit method using space-based missions \citep{borucki2010kepler, batalha2013planetary, burke2014planetary, rowe2015planetary, mullally2015planetary, coughlin2016planetary, morton2016false}, and confirmed as exoplanets with radial velocity measurements, the analysis of transit timing variations, and/or through statistical validation \citep{thompson:2018}.

Kepler's successor, NASA's Transiting Exoplanet Survey Satellite \citep[TESS;][]{ricker2014transiting} that was launched in April 2018, and a future mission,  The PLAnetary Transits and Oscillations of stars \citep[PLATO;][]{rauer2014plato} mission, are telescopes whose main objective is the detection of exoplanet candidates. TESS is monitoring millions of stars with high photometric precision to detect signals of transiting extrasolar planets. Unlike Kepler, TESS is surveying a large fraction of the sky (approximately 90\% of it), searching for planets around the nearest and brightest stars. The size of each TESS sector is approximately 2300 square degrees, whereas Kepler had a field of view of approximately 115 square degrees. TESS observes a sector for 27 days, and then the spacecraft is reoriented to observe the next sector. The main objective of TESS is to find hundreds of planets in transit smaller than Neptune, including dozens that are comparable in size to Earth and amenable for atmospheric characterization \citep{Ricker:2015}. \par

The identification of transiting exoplanet candidates from photometric time series typically consists of a two-step process. The first step involves the removal of long-term flux variations of stellar and/or instrumental origin, followed by the identification of periodic transit-like signals by phase folding the light curve to different orbital periods, times of transit, and transit durations \citep[e.g.][]{tenenbaum:2012}. The products of this process are known as Threshold Crossing Events (TCEs), which can be exoplanet candidates but also false positives. Some phenomena can mimic a transit signal; among these impostors are eclipsing binaries, instrumental artifacts, noise, and stellar variability. The second step for identifying exoplanet candidates consists of the vetting of the TCEs, by rejecting false positives based on the properties of the phase-folded transiting light curves, stellar properties, and/or follow-up observations. This process is labor intensive and sensitive to human error, and increasingly requires a significant amount of effort since there are large volumes of data to be examined. Thus, identifying exoplanet transit signals has become an increasingly data-science driven task. In response to this challenge, several efforts have been made to classify candidates automatically from space-based and ground-based surveys. \par

In the era of DL \citep[]{lecun2015deep, sun2017revisiting}, neural network (NN) architectures are applicable in almost all areas of scientific research, producing great advances in image processing and classification, facial recognition, voice recognition and Natural Language Processing (NLP) \citep{krizhevsky2017imagenet, ciregan2012multi, he2016deep, voulodimos2018deep, vaswani2017attention}. Furthermore, the area of DL and many areas in astronomy largely share the philosophy of data-driven study and analysis. In particular, DL  techniques have been applied for the search and examination of exoplanet candidates. One of the most outstanding applications in this area has been the use of CNNs, firstly explored by \citet{pearson2018}, \citet{Shallue_2018} and \citet{schanche2019machine}. 
These approaches developed CNNs trained with data from Kepler and from the Wide Angle Search for Planets \citep[WASP;][]{pollacco2006wasp} respectively, achieving successful results in the identification of exoplanetary signals. Furthermore, they have demonstrated that the incorporation of additional information to a CNN model, e.g., the information of the flux centroid pixel position, considerably improves their performance \citep[Exonet;][]{ansdell2018scientific, osborn2020rapid}.  The centroids represent the pixel position of the centre of the light. This information allows us to know the location of the transits in the pixels. In general, experts perform this analysis to verify whether the transit signals are the product of sources other than an exoplanetary transit signal \citep{Shallue_2018}. \par


Even though current approaches have reached a high level of accuracy, the results are not interpretable, i.e., it is not possible to understand why the network classifies a given system as a transiting planet or as a false positive. Interpreting the predictions of a model can lead to a better understanding of the model and to a more efficient use of its outputs. Furthermore, interpreting the result of a model prediction generates a greater value by knowing the importance of certain features, in such a way that they can be later justified by experts. This could help future prediction analysis in differentiating the characteristic features of a planetary transit signal from those of a false positive. Unfortunately, the explainability of models such as CNNs is low because convolutional filters are difficult to understand, limiting the interpretability of the hidden representation in the network \citep{zhang2019interpreting, vilone2020explainable}.\par


The application of one-dimensional CNNs on time series data has had good results. However, when working with this type of data, CNNs require many layers to capture dependencies \citet{hawkins2004problem, lakew2018comparison}. Typically a layer is a container that receives a weighted input. In fact, building models capable of learning long-range dependencies is challenging for the majority of sequential modeling tasks. In the case of classification of transiting exoplanet signals, the state-of-the-art DL approaches propose a global and a local view of the transit signal as input to the CNN model \citep{Shallue_2018}. The light curve of the ``global view'' shows the complete phase-folded light curve, i.e., the light curve over an entire orbital period. The ``local view'' is centered around the transit. Regarding the length of both sequences, the length of the global view has more than twice the number of observations than the local view. In that sense, the NNs for the global view of a planetary transit have many more layers compared to the local view. This is because it is a longer sequence and requires a greater network depth to capture dependencies. Although the layers of a CNN work in parallel, the length and dimensionality of the sequential data could generate some computational limitations in this type of network \citep{vaswani2017attention}. \par

Attention mechanisms were proposed by the NLP community to overcome the computational problems of CNNs and Recurrent neural networks \citep[RNNs;][]{cho2014learning}. The Transformer architecture, which is mainly based on self-attention mechanisms \citep{vaswani2017attention}, have gained popularity for its performance on sequentially structured data. This type of architecture is capable of working well in complex scenarios for other kinds of models (RNN, CNN), such as sequences \citep{hawkins2004problem, lakew2018comparison, karita2019comparative}. In general, time series data have more long-range correlated information. For example, a planetary transit signal depends on duration and period, which causes a correlation in light curve observations. In this sense, attention mechanisms can capture this information better. \par



In this paper, we propose a new deep neural network based on the Transformer architecture for the automatic classification of exoplanet transit signals. Unlike previous DL models for recognizing exoplanet signals, we provide a context and interpretability of the model output that would help to better understand the patterns of each signal. Thus, we propose a model that provides that interpretability when classifying a signal as a exoplanet transit or as a false positive signal, in turn, showing the characteristics that differentiate both classes. We include the measurements of the flux centroid pixel positions to improve the representation of an exoplanetary transit signal. Furthermore, we show the relevance of using centroids in terms of interpretability and how it affects the final prediction of the model. We show that the input representation is essential for a model to understand the importance of specific characteristics of a transit signal in a light curve. We also take into account that the transits have temporal information (orbital period, transit duration). Specifically, for the order of the sequence of observations, we injected time information from the light curves into the positional encoding of the input representations. Then, we generate a more representative input of the flux and its centroid pixel position. In addition to the information on the centroids, we include information on the parameters of the transit and the host star like \citet{ansdell2018scientific}, \citet{osborn2020rapid} and \citet{rao2021nigraha}. Finally, we identify the most important parameters to differentiate an exoplanet transit from a false positive according to the self-attention mechanism.


This paper is organized as follows. Section~\ref{sec:background} introduces the main concepts that inspired this work, where we briefly describe the self-attention mechanisms related to the original architecture of the Transformer \citep{vaswani2017attention} in order to understand our proposal for planetary transit signal recognition.   Section~\ref{sec:relatedwork}, gives an account of previous work, which is divided into two parts. First, we present a review of machine learning models for exoplanetary transit signal recognition. In the second part of the review, we show how the attention mechanism has been applied in astronomy. Section~\ref{sec:ourmethod} explains the proposed model in detail. Section~\ref{sec:datasets} describes the data sets used. The training process and performance evaluation are described in Section ~\ref{sec:training}. Section~\ref{sec:results} shows our classification result obtained using our model. Finally, in Section~\ref{sec:conclusions} we present overall conclusions and future work.

\section{Background Theory}
\label{sec:background}
In this section, we briefly describe the main components and concepts to understand our method. First, in Section~\ref{sec:background_attention}, we introduce the attention mechanism. Then, in Section~\ref{sec:self_attention} we give way to the description of the principal core of the Transformer architecture \citep{vaswani2017attention}, which is the self-attention mechanism, explaining, in turn, each component of this architecture that inspired this work. Lastly, in Section \ref{sec:positional_encoding}, we describe the positional assignment mechanism of each element in a sequence.

\subsection{Attention mechanisms}
\label{sec:background_attention}

DL, a typical branch of machine learning methods, is a set of computational heuristics to train parametric functions structured in a layered form to perform a certain task \citep{lecun2015deep, goodfellow2016deep}, and that has been successful in classification problems from a variety of application domains. The architecture relies on several layers of NNs of simple interconnected units and uses layers to build increasingly complex and useful features by means of linear and non-linear transformation. This family of models is capable of generating increasingly high-level representations \citep{lecun2015deep}. 


Attention theory is one of the most prominent ideas in the DL community, which emerged from the applications of NLP. Typically, NNs are inspired by different actions and processes of the human brain. Consequently, the attention mechanism also attempts to mimic the action of selectively focusing on some relevant things while ignoring others. Human beings tend to pay attention to ideas, things, language, etc., when carrying out specific tasks or solving problems. In NLP, the attention mechanism emerged as an improvement over the sequence modelling encoder-decoder-based architecture for the translation system \cite{bahdanau2014neural}.


In astronomy, light curves are sequences as a function of time, where each observation in the sequence is related to the observations that precede or follow it. For example, in the case of light curves of transiting planets these sequences have observations based on an order according to orbital period and transit duration, similar to the structure of sentences in a paragraph. In NLP, each word is an element of the sequence. In light curves, each observation represents an element of the sequence. So, through the attention mechanism, we can selectively focus on some observations that characterize an astronomical event while ignoring others, for example, observations that do not contain any relevant event. This mechanism and its variants have been used not only in modeling sequences but also in computer vision \citep{dosovitskiy2020image, wu2020visual, li2020behrt}, speech processing \citep{karita2019comparative, dong2018speech} and other areas within DL \citep{han2020survey, huang2018music}. \par

The traditional attention mechanism was designed to connect the encoder with the decoder, feeding the decoder with information on each hidden state of the encoder. This piece allows comparing the input and output of the phrase, an image, or a previous step. With sequence modeling, it is possible to build models that take as input a vector of symbols, also named vector embedding (e.g., representation of words, images, numbers, etc.) $(x_1, x_2, ..., x_n)$, where $n$ is the sequence length. The encoder transforms this input into a continuous vector $z$, i.e., vectorizes the input sequence. Then, this is the input to the decoder, which attends to the encoded representation and produces the output sequence $(y_1, y_2, ...,y_n)$ of symbols. The output type of the prediction depends on the specific task. For example, the output type is a discrete scalar for binary or multi-class classification. In the case of a regression task, this output is a continuous value. \par

This mechanism contrasts with other architectures, such as RNNs, which process data sequentially. In other words, the RNNs process element by element to access the content of the last input element of the time series. 
This type of process tends to make the network forget the information from positions that are further away or, in certain instances, merge the information from adjacent positions across the time series \citep{lakew2018comparison}. In the case of CNN's, they learn by batch with parallelism across layers, being fast and effective in capturing dependencies across the time series. However, in long sequences, capturing the dependencies between different combinations of input elements requires deeper NNs, which can be cumbersome and impractical \citep{hawkins2004problem}.

\begin{figure*}
\vspace{5mm} 
\begin{center}
\includegraphics[width=0.9\textwidth]{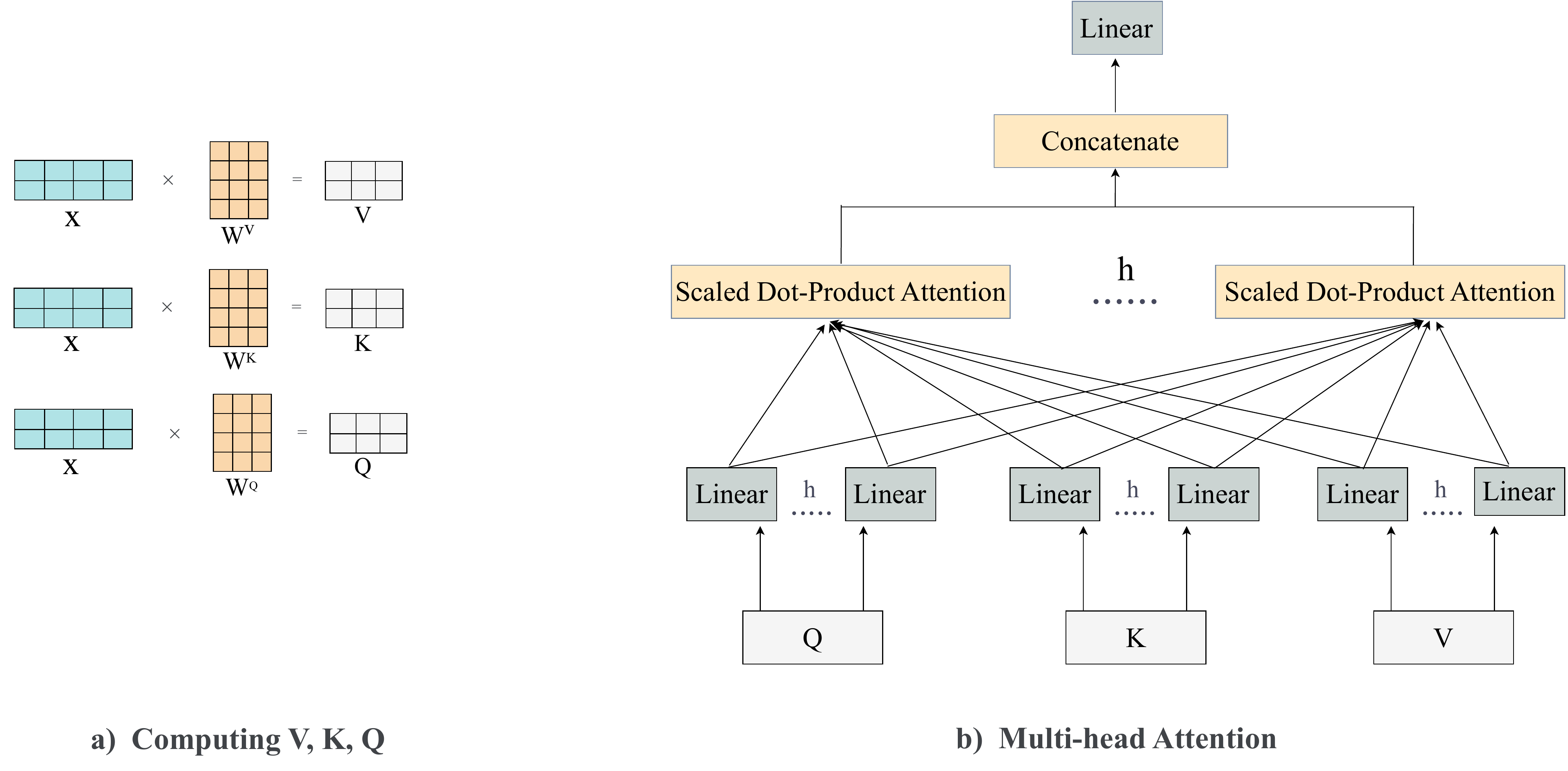}
\vspace{2mm} 
\caption{\label{fig:multi_head_attention}
Description of multihead attention mechanism defined in \citep{vaswani2017attention}. a) Multiplication of the input sequence $X$ by the weight matrix $\mathbf{W}^V$, $\mathbf{W}^K$ and $\mathbf{W}^Q$ to produce $\mathbf{V}$, $\mathbf{K}$ and $\mathbf{Q}$ respectively. b) Multihead attention, where values $\mathbf{V}$, keys $\mathbf{K}$, and queries $\mathbf{Q}$ are linearly projected $h$ times with different learned linear projections. Then, multihead attention component generates $h$ dimensional output values in parallel. Finally, these are concatenated and projected, resulting in the final values.
}
\vspace{2mm}
\end{center}
\end{figure*}

\subsection{Self-attention}
\label{sec:self_attention}



From the theory of attention mechanism, a new form of attention emerged, called self-attention, and with it, a new class of models, which are mainly focused on sequential data. In contrast to the traditional attention mechanism, which attends to different parts of the input or output sequences while making the predictions, the self-attention mechanism allows each element of the input sequence to interact with each other while encoding each of the input elements. \par


Self-attention, also referred as intra-attention is designed to find important relationships between each input element \citep{vaswani2017attention}. This allows the network to find which elements it should ``pay attention'' to, similar to the way a human being understands words in a sentence; for example, it complements the subject of a sentence with a verb or adjective. Based on this idea, in light curve time series, the self-attention allows the model to learn the correlation between the current observations and the other observations in the same sequence. For example, a transit signal contains observations, such as noise, some star variation, or brightness dimming, which could be highly correlated with other values in the same sequence. This information could be essential to determine if a signal belongs to an exoplanet. Furthermore, the self-attention quantifies the relationships between the observations without the condition that the operations follow a sequential order. This is beneficial so that operations can be executed in parallel.



The aim of self-attention is to compute the similarities of each input within the sequence with each other. It is computed using three vectors, $Q$, $K$ and $V$, referred to as query, key, and value, respectively. Vectors $Q$ and $K$ are compared to calculate similarities and obtain scores/weights for vector $V$. To compute these vectors as three different subspace representations, the input sequence is projected using three matrices $\mathbf{W}^Q$, $\mathbf{W}^K$ and $\mathbf{W}^V$, which are initialized weight matrices with independent random values with mean 0 and variance 1 by definition. Given a light curve, which is the input sequence and is defined as $X \in \mathbb{R}^{n \times d}$, where $n$ is the number of observations of the input sequence, and $d$ is the input dimensionality and the three weight matrices $\mathbf{W}^Q$, $\mathbf{W}^K$ and $\mathbf{W}^V$, the $Q$, $K$ and $V$ are computed as:
. \par





\begin{equation}
\begin{split}
\mathbf{Q=XW^{\mathnormal{Q}} \in \mathbb{R}^{\mathnormal{n \times d_q}}},\\
\mathbf{K=XW^{\mathnormal{K}} \in \mathbb{R}^{\mathnormal{n \times d_k}}},\\
\mathbf{V=XW^{\mathnormal{V}} \in \mathbb{R}^{\mathnormal{n \times d_v}}}.
\label{eq:attention_qkv}
\end{split}
\end{equation}

Figure~\ref{fig:multi_head_attention}(a) shows this process, where the first step is multiplying each of the encoder input sequence $X$ by three weights matrices ($\mathbf{W}^Q$, $\mathbf{W}^K$ and $\mathbf{W}^V$) that we trained during the training process. We then take the $\mathbf{Q}$, $\mathbf{K}$ and $\mathbf{V}$ matrices, and supply them as input to the attention function. A self-attention matrix $\mathbf{A} \in \mathbb{R}^{n \times d_v}$ is often referred to as the scaled dot-product attention and can be written as:

\begin{equation}
\mathbf{A}=\mathrm{Attention\mathbf{(Q,K,V)} = softmax \left ( \frac{\mathbf{QK}^\top}{\sqrt{d_k}} \right ) \mathbf{V}}
\label{eq:scaled-dot-product}
\end{equation}
The attention layer takes the input in the form of those three parameters (after positional encoding). The three parameters have similar structure, where each element in the sequence is represented by a vector. \par

\subsubsection{Multihead attention}
In the Transformer architecture, the attention module consists of several self-attention heads running in parallel. This module is also called multihead attention, which allows the neural network to learn from $h$ different representation subspaces of $Q$, $K$, and $V$ simultaneously. This process, in turn, generates a richer representation for the model. \par


In multihead attention (see Figure~\ref{fig:multi_head_attention}(b)), the key, values, and queries are linearly projected $h$ times, i.e., we can repeat its computations multiple times in parallel \citep{vaswani2017attention}. Then, those results are concatenated, and we obtain the final values:

\begin{equation}
\label{multihead_eq}
\mathrm{MultiHead\mathbf{(Q,K,V)} = 
\mathbf{W}^O \begin{bmatrix}
 h_1\\
.\\
.\\
.\\
 h_h
\end{bmatrix}
}
\end{equation}
with learned output weights $\mathbf{W}^O \in \mathbb{R} ^{hd_v\times d} $, and where:
\begin{equation}
h_i = \mathrm{Attention(\mathbf{XW}^{\mathnormal{Q}}_i, \mathbf{XW}^{\mathnormal{K}}_i, \mathbf{XW}^{\mathnormal{V}}_i) }
\end{equation}
with $h_i \in \mathbb{R}^{n \times d_v}$. Then, the final result of this linear transformation, Equation~(\ref{multihead_eq}), is the multi-headed attention matrix with dimension $n \times d$. In an attention with $h=1$, the network learns from a single behavior of the information. On the contrary, the use of multiple heads of attention allows the network to learn from a final representation richer in information since each ``head'' of attention is potentially focusing on different parts of the input sequence. To determine the appropriate number of heads, we must analyse how the model captures the dependencies along the sequence. For this, we analysed the individual attention of each head through attention entropy. In this way, we can observe how attention weights are concentrated on specific observations (see Section~\ref{individual_attention_heads}). Choosing the number of heads is analogous to the number of kernels in a convolutional layer. 







\subsection{Positional encoding}
\label{sec:positional_encoding}
The Transformer architecture avoids sequential operations in favor of a multihead self-attention mechanism with parallel computation. The positional encoding describes the location or position of a token in a sequence, such that each position is assigned to a unique representation. The idea of injecting a position to the sequence was first introduced in \citet{gehring2017convolutional}, where it was used in the context of sequence modeling with CNNs. In the original architecture of the Transformer, the process of positional embedding is a new way of representing the values of a word. \citet{vaswani2017attention} implement a positional encoding from a vector of indices and adds a trigonometric function:

\begin{equation}
{ \mathbf{PE_{\mathnormal{t, i}}} :=}
\left\{\begin{matrix}
 \mathrm{sin}(p,\omega_i), & \mathrm{if }\; p = 2i   \\
 \mathrm{cos}(p,\omega_i), & \mathrm{\; \; \; \; \; if} \; p = 2i+1 \\
\end{matrix}\right.
\label{equation_pe}
\end{equation}
where $p$ represents the position of an object in the input sequence, $i$ that is used for mapping to indices of the positional encoding and $\omega_i$ represents the frequency of progression, which is computed as:
\begin{equation}
\omega_i = \frac{1}{10000^{2i/d}}.
\end{equation}
where $d$ is the dimension of the output embedding space.



\section{RELATED WORK}
\label{sec:relatedwork}
In this section, we describe the previous works related to our proposal, which are divided into two parts. In Section~\ref{sec:relatedwork_exoplanet}, we present a review of approaches based on machine learning techniques for the detection of planetary transit signals. Section~\ref{sec:relatedwork_attention} provides an account of the approaches based on attention mechanisms applied in Astronomy.\par

\subsection{Exoplanet detection}
\label{sec:relatedwork_exoplanet}
Machine learning methods have achieved great performance for the automatic selection of exoplanet transit signals. One of the earliest applications of machine learning is a model named Autovetter \citep{MCcauliff}, which is a random forest (RF) model based on characteristics derived from Kepler pipeline statistics to classify exoplanet and false positive signals. Then, other studies emerged that also used supervised learning. \cite{mislis2016sidra} also used a RF, but unlike the work by \citet{MCcauliff}, they used simulated light curves and a box least square \citep[BLS;][]{kovacs2002box}-based periodogram to search for transiting exoplanets. \citet{thompson2015machine} proposed a k-nearest neighbors model for Kepler data to determine if a given signal has similarity to known transits. Unsupervised learning techniques were also applied, such as self-organizing maps (SOM), proposed \citet{armstrong2016transit}; which implements an architecture to segment similar light curves. In the same way, \citet{armstrong2018automatic} developed a combination of supervised and unsupervised learning, including RF and SOM models. In general, these approaches require a previous phase of feature engineering for each light curve. \par


The application of DL for exoplanetary signal detection has evolved rapidly in recent years and has become very popular in planetary science.  \citet{pearson2018} and \citet{zucker2018shallow} developed CNN-based algorithms that learn from synthetic data to search for exoplanets. Perhaps one of the most successful applications of the DL models in transit detection was that of \citet{Shallue_2018}; who, in collaboration with Google, proposed a CNN named AstroNet that recognizes exoplanet signals in real data from Kepler. AstroNet uses the training set of labelled TCEs from the Autovetter planet candidate catalog of Q1–Q17 data release 24 (DR24) of the Kepler mission \citep{catanzarite2015autovetter}. AstroNet analyses the data in two views: a ``global view'', and ``local view'' \citep{Shallue_2018}. \par



Based on AstroNet, researchers have modified the original AstroNet model to rank candidates from different surveys, specifically for Kepler and TESS missions. \citet{ansdell2018scientific} developed a CNN trained on Kepler data, and included for the first time the information on the centroids, showing that the model improves performance considerably. Then, \citet{osborn2020rapid} and \citet{yu2019identifying} also included the centroids information, but in addition, \citet{osborn2020rapid} included information of the stellar and transit parameters. Finally, \citet{rao2021nigraha} proposed a pipeline that includes a new ``half-phase'' view of the transit signal. This half-phase view represents a transit view with a different time and phase. The purpose of this view is to recover any possible secondary eclipse (the object hiding behind the disk of the primary star).

%

\subsection{Attention mechanisms in Astronomy}
\label{sec:relatedwork_attention}
Despite the remarkable success of attention mechanisms in sequential data, few papers have exploited their advantages in astronomy. In particular, there are no models based on attention mechanisms for detecting planets. Below we present a summary of the main applications of this modeling approach to astronomy, based on two points of view; performance and interpretability of the model.\par

The application of attention mechanisms has shown improvements in the performance of some regression and classification tasks compared to previous approaches. One of the first implementations of the attention mechanism was to find gravitational lenses proposed by \citet{thuruthipilly2021finding}. They designed 21 self-attention-based encoder models, where each model was trained separately with 18,000 simulated images, demonstrating that the model based on the Transformer has a better performance and uses fewer trainable parameters compared to CNN. A novel application was proposed by \citet{lin2021galaxy} for the morphological classification of galaxies, who used an architecture derived from the Transformer, named Vision Transformer (VIT) \citep{dosovitskiy2020image}. \citet{lin2021galaxy} demonstrated competitive results compared to CNNs. Another application with successful results was proposed by \citet{zerveas2021transformer}; which first proposed a transformer-based framework for learning unsupervised representations of multivariate time series. Their methodology takes advantage of unlabeled data to train an encoder and extract dense vector representations of time series. Subsequently, they evaluate the model for regression and classification tasks, demonstrating better performance than other state-of-the-art supervised methods, even with data sets with limited samples.

Regarding the interpretability of the model, a recent contribution that analyses the attention maps was presented by \citet{bowles20212}, which explored the use of group-equivariant self-attention for radio astronomy classification. Compared to other approaches, this model analysed the attention maps of the predictions and showed that the mechanism extracts the brightest spots and jets of the radio source more clearly. This indicates that attention maps for prediction interpretation could help experts see patterns that the human eye often misses. \par

In the field of variable stars, \citet{allam2021paying} employed the mechanism for classifying multivariate time series in variable stars. And additionally, \citet{allam2021paying} showed that the activation weights are accommodated according to the variation in brightness of the star, achieving a more interpretable model. And finally, related to the TESS telescope, \citet{morvan2022don} proposed a model that removes the noise from the light curves through the distribution of attention weights. \citet{morvan2022don} showed that the use of the attention mechanism is excellent for removing noise and outliers in time series datasets compared with others approaches. In addition, the use of attention maps allowed them to show the representations learned from the model. \par

Recent attention mechanism approaches in astronomy demonstrate comparable results with earlier approaches, such as CNNs. At the same time, they offer interpretability of their results, which allows a post-prediction analysis. \par

\section{Method Description}
\label{sec:ourmethod}

In this section, we describe our methodology, which is able to identify if a transit signal belongs to an exoplanet or a false positive. Our work is mainly motivated by the classification of transiting exoplanet candidates originated from the TESS mission. The proposed architecture considers the exoplanetary transit signal, centroid information, and stellar and transit parameters. Our method consists of three main steps, which we describe in this section. First, in Section~\ref{sec:preprocessing}, we explain how the input representation is constructed. Then, Section~\ref{sec:proposed_model} shows the design of our proposed model. Finally, Section~\ref{sec:attention_entropy_model} describes the methodology we adopted to analyse the resulting attention matrices. \par

\subsection{Data preprocessing}
\label{sec:preprocessing}

A light curve is defined by flux observations taken over a given time. We process light curves belonging to TCEs from various TESS sectors \footnote{https://archive.stsci.edu/tess/bulk\_downloads.html\#catalogs}. In the case of transit signals, we consider the ``local'' and ``global'' views (similar to \citet{Shallue_2018}), both of which depend on the duration of the transit. Then, we include the information of the centroids in the input representation, and we observe the behavior of the attention weights on the main characteristics that differentiate a planetary transit from a false positive. Additionally, we include information on the stellar and transit parameters, and we show the importance of each parameter to distinguish a transit. \par

\subsubsection{Flux and centroid time series}
\label{sec:preprocessing_flux}

We perform some additional steps to prepare the light curves with a method similar to \citet{Shallue_2018}, \citet{ansdell2018scientific}, \citet{osborn2020rapid}, \citet{yu2019identifying} and \citet{rao2021nigraha}. We use the Pre-Search Data Conditioning Simple Photometry flux (PDCSAP\_FLUX) that has the instrumental systematic removed. Within the processing of the light curves, we remove outliers if their flux values are at least $5\sigma$ above the median of the light curve. Then, we remove the low frequency trend of light curve applying the Savitzky–Golay filter using the {\tt\small flatten()} function from the Lightkurve toolkit \citep{2018ascl.soft12013L}. Then, we process each light curve to bin the phase-folded local and global views similar to \citet{Shallue_2018}. For all processing procedures, we use the {\tt\small Lightkurve}\footnote{https://docs.lightkurve.org/index.html} and {\tt\small Astropy API}{\footnote{astropy.stats.sigma\_clipping}}. For phase-folding the local and global views, we take the transit ephemeris available in the ExoFOP-TESS\footnote{https://exofop.ipac.caltech.edu/tess/} catalogue. For the local view, which depends on transit duration $t_{dur}$, we zoomed in on the transit between $-2.0\cdot$ $t_{dur}$ to $2.0\cdot$ $t_{dur}$. In the case of the global view, it corresponds to the full view of the light curve with a phase between $-0.5$ to $0.5$. In addition to the PDCSAP\_FLUX, we included the centroid information into our input. Incorporating this information was applied by \citet{ansdell2018scientific, osborn2020rapid}. We used the CCD row and column position information of the target centroid using a Point Spread Function (PSF) model that is available in the PDC files from the TESS Input Catalog. And if this information is not available in some files, we used the CCD row and column position of the flux-weighted centroid of the target (MOM centroids), that are less robust against background noise compared to centroids using a PSF model. Having data originated from different distributions or samples is very common in machine learning, and one way to transform the data to comparable scales is by drawing on standardization techniques. Based on this, we standardize the values by removing the mean and scaling to unit variance for both centroids. \par

\subsubsection{Stellar and transit parameters}
\citet{ansdell2018scientific} show that providing stellar and transit parameters can improve the model performance for the recognition of a planetary signal. Unlike \citet{osborn2020rapid} and \citet{rao2021nigraha}, we prefer to take only some stellar and transit parameters since, in many cases, much of the information was not available or had null values. The parameters we process are the metallicity [Fe/H], temperature $T_{eff}$ and magnitude $TESS_{Mag}$ of the host star. Regarding the transit parameters, we consider the period, duration and depth, which were taken from the TCE statistics. \par

\subsection{Model}
\label{sec:proposed_model}

The core of our architecture is the self-attention mechanism, but it also contains other important components. These components are: i) input embedding, which consists of the representation of the input, ii) positional encoding, iii) multihead self-attention, iv) pooling layers, which are designed to greatly reduce the number of input features. This process is called downsampling, v) linear layer, also known as fully-connected layers, in which each neuron applies a linear transformation to the input vector. 

\begin{figure*}
\begin{center}
\vspace{5mm}
\includegraphics[width=0.99\textwidth]{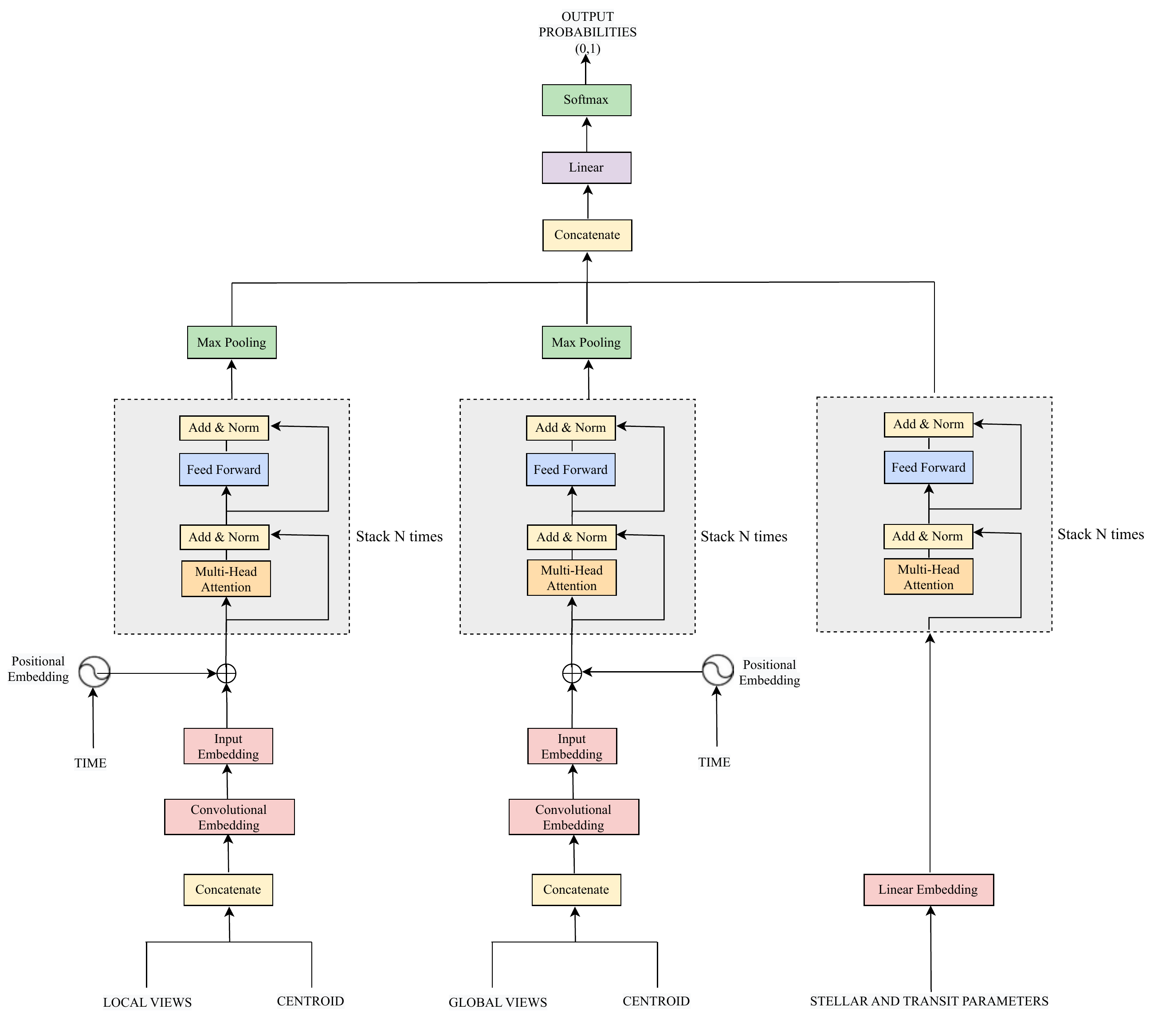}
\vspace{-2mm} 
\caption{\label{fig:model_att}
Scheme of the proposed architecture for the analysis of transit signals. The architecture consists of three encoders, where each encoder is designed for i) local view, ii) global view, and iii) stellar and transit parameters. The first encoder block corresponds to the representation of the local view. The input is the concatenated time series of flux time and centroid. A convolutional embedding layer follows, which transforms this input into a vector representation with a positional encoding. This vector is the input of the multihead attention block, which is comprised of $N$ identical layers and a feed-forward sub-layer. The output of the multihead attention block is passed as input to a max pooling layer. The second encoder is implemented for the global view. Then, both time series are concatenated and follow the same steps as the first encoder. The third encoder contains the stellar/transit parameter information and is followed by a linear layer to transform this information into a vector representation. This representation is the input of the multihead attention block. Finally, the output of each encoder is passed as input to a linear layer with softmax for class prediction.}
\vspace{-1mm}
\end{center}
\end{figure*}

Our proposed architecture is shown in Figure~\ref{fig:model_att}, which contains the necessary encoder modules to incorporate the flux, centroid information (local and global view), stellar/transit parameters, and time of folded light curves with a particular period. In contrast to the original architecture by \citet{vaswani2017attention}, we consider the decoder as the module with the activation function for a binary classification. Additionally, we implement three encoders. The first encoder is for the local view light curve, the second one for the global view, and the third one for the stellar/transit parameters. We transform the information of the light curves (local and global view) into a vector representation through a convolutional layer. A positional encoding is added to this new representation, this representation is the input to the multihead attention block. The output of this block is then passed through a new max pooling layer. In the case of the third encoder, a linear layer transforms the input that corresponds to the stellar and transit parameters, and is the input to the multihead attention block. Finally, we concatenate each output of each block and pass them as input to a linear layer with softmax for class prediction. In the rest of this section, we describe each of the components of our architecture in detail.

\subsubsection{Input embedding}
\label{sec:input_embedding_sec}

We transform the information of the light curves (local and global view) into a vector embedding through a convolutional layer.

The first step for each piece of information is to generate an input sequence before building the input embedding of each encoder in our model. In the case of the ``local view'', after the phase folding process, we generate a sequence of observations in terms of the flux and centroid $x_1, x_2, x_3, ..., x_n$ and $c_1,c_2, c_3, ..., c_n$ respectively. Then, we concatenate both time series to create a sequence as input of the first vector embedding. In this way, we obtain a single sequence for the ``local view'' $\{x'_1,x'_2, ..., x'_n\}$ which is equivalent to $\{(x_1,c_1),(x_2,c_2), ..., (x_n,c_n)\}$. For the input sequence of the ``global view'', we perform the same procedure as for the local view. \par




As a second step, we build our input data representation. In the context of DL, the data representation is also called vector embeddings and consists of a representation of symbols from data (e.g., words, observations, images, etc.) in a transformed space. Embeddings can be vectors of 1s and 0s, numeric, or neural networks (NNs). In general, an embedding captures dependencies of the input across different variables in the embedding space; this generates a more expressive input vector for the network and facilitates sequential modeling. To obtain our input embedding, we apply a 1D convolutional layer to our input sequence. We chose to use this type of representation since the CNN of 1D can identify local patterns or features within the convolution window. Each feature in our input sequence only needs to be learned once and a pattern learned at one position can also be recognized at a different position. Thus, given a sequence of observations in terms of the flux and centroid $\{x'_1,x'_2, ..., x'_n\}$, $x' \in \mathbb{R}^m$, where $m$ is the number of variables of our time series, we apply 1D convolution to the sequence. We denote the convolution filter coefficients as $\mathbf{w} \in \mathbb{R}^{n \times k}$, and our input embedding as $\mathbf{w}^{\left ( k \right )}_i \cdot x^{\left ( k \right )}_{i-1}$ for each $x'_i$ at position i. For 1D, the kernel $k=1$ slides along one dimension transforming an m-dimensional space to a $d$-dimensional space for every $t$.





\subsubsection{Positional embedding}

Once the embedding vector is built, we add an encoding that assigns the position of each element of the embedding vector. This encoding is added to the local and global view (see Figure~\ref{fig:model_att}). \citet{vaswani2017attention} implement a positional encoding from a vector of indices and add a trigonometric function as described in Section~\ref{sec:positional_encoding}. That implementation is applied to sentences of words, where the space of a word with its predecessor is the same. In the case of light curves, the time series are irregular, i.e., a sequence of observation time and value $(t_n, x_n)$. In the observations we analysed of the TESS telescope, the cadence is $\sim$2min, which means that it is mostly constant but with some interruptions. To account for these slight variations, we inject the observation time $t$ of the light curve instead of the position $p$ with the same trigonometric function defined by \cite{vaswani2017attention} (see Equation~\ref{equation_pe}). Thus, our final positional encoding is $\mathbf{P} \in \mathbb{R}^{n \times d}$ (one for each element of the sequence), which is a $d$-dimensional vector that contains information about the specific position in a time series of a light curve of our input embedding. Additionally, we also experimented with injecting $p$ as done initially by \cite{vaswani2017attention} instead of $t$, and as a result, we found that both encodings produce almost identical results (see Section ~\ref{ablation_study}). This is due to the cadence (mostly constant) of the TESS observations. These results would probably not be the same for light curves with larger or random cadences, which usually occur with ground-based telescopes.\par


%


\subsubsection{Attention module}

In the attention module (gray blocks in Figure~\ref{fig:model_att}), we consider the self-attention mechanism described in Section~\ref{sec:self_attention}. This mechanism is the core of every encoder block. Our input embedding is divided into multiple heads, so the split sections of the embedding learn different aspects of the meanings of each observation as it relates to other observations of the same sequence. With this process, the model captures the most important features of the sequence.

In the original self-attention encoder by \citet{vaswani2017attention}, the mask is used to identify which tokens or values are padding from being part of the attention score; i.e., with the mask, it is possible to control the interaction between each input observation with other input observations. In our model, we do not mask the sequence, i.e., the mechanism observes past and future values. In this way, our model observes when the brightness of the star begins to decrease, and in parallel is able to observe when its brightness increases. With this, we get the mechanism to compare the beginning and ending of a transit signal.


\subsubsection{Max pooling layer}
The output of each encoder enters a layer of pooling. Pooling layers provide a down sampling approach that reduces features into patches. We introduce a max pooling layer that is added after each implemented encoder for the global and local views. As mentioned before, this reduces the sample of the entities of the next layer, allowing the features in the next layer to have a broader view of the input.

 
\subsubsection{Final Linear and Softmax layer}
The Linear layer is a simple fully connected neural network that projects the vector produced by the encoder, into vector called a logits vector. We consider the decoder as our activation function for binary classification, and we use a softmax layer for this task, which returns a vector of probability distribution $z$. The label of the predicted class is the one with the highest probability score. The loss function is defined by the equation:
\begin{equation}
\mathcal{L} = \mathrm{-\frac{1}{n} \sum_{i=1}^{n} (y_i \cdot \mathrm{log} (\hat{y}_i) + (1-y) \cdot \mathrm{log} (1-\hat{y}_i) ) }
\end{equation}
$y_1, y_2, ..., y_n$ are the true labels of all samples in the training set (defined to be either 0 or 1), and $\hat{y}_i$ is the prediction probability of the model.

\subsection{Analysis of the attention heads}
Part of our proposed methodology consists of analysing the attention distributions that the model captured. For this, we measure the entropy in each attention head and average the attention weights for each time step. In this section, we describe both criteria.

\subsubsection{Attention entropy}
\label{sec:attention_entropy_model}
We analysed individual attention heads to observe the patterns of transit signals that have been learned. First, we measure whether the attention is focused on all observations of the phase-folded light curve or only when the transit occurs. In other words, we measure whether attention is focused when the star's brightness decreases. One way to measure this is by computing the average entropy of attention probabilities for each layer as a ``sparsity measure'' of attention. We compute the attention entropy (\ref{entropy_eq}) \citep{ghader2017does} to measure how the attention distribution is focused at the timestep $t$:

\begin{equation}
\label{entropy_eq}
\mathrm{\mathbf{E_{\mathrm{At}_t}} =  - \sum_i^{\left| x\right|} At(x_i, y) \log At(x_i, y) }
\end{equation}
where $x_i$ represents the $i-$th source token (observation), $y$ denotes the prediction label, and $At_t(x_i, y)$ is the attention distribution at timestep $t$. Then, we average the entropy of attention over all timesteps to get the final entropy.

\subsubsection{Attention weights }
To interpret the attention distribution, we averaged the attention weights that the model attended along the observation sequence of dimension $n$ for an element at time $t$:

\begin{equation}
\label{attention_weights}
\mathrm{At(x_t) =  \frac{1}{n}\sum_i^n At(x_i) }, \; \; \; \mathrm{for}\; \mathrm{a} \; \mathrm{time} \;\; t
\end{equation}

\subsection{Complexity}

The complexity of the proposed approach for each layer is given by $O(n^2 \cdot d)$, where $n$ is the sequence length ($n=100$ for the local view and $n=200$ for the global view of the transit signal), $d = 512$ and as a result $n\ll d$. In the case of CNN's, layers are generally more computationally expensive, by a factor of $k$, which is a filter to extract the features from the light curve. The complexity per each CNN layer is given by $O(k \cdot n \cdot d^2)$ \citep{vaswani2017attention}.  So, the increase of layers in the network generates a higher complexity that could lead to an overfitting of the model by acting as a filter \citep{hawkins2004problem}. As a result, our approach is faster than a CNN.\par



\section{DATASETS}
\label{datasets}
\label{sec:datasets}


We used the pre-processed TESS light curves from the Mikulski Archive for Space Telescopes (MAST) \footnote{https://archive.stsci.edu/tess/bulk\_downloads/bulk\_downloads\_ffi-tp-lc-dv.html}, which have 2-minute cadence observations of 200,000 to 400,000 selected stars. We use two datasets to collect the light curves. The first dataset corresponds to the TESS TOI catalog (ExoFOP-TESS)\footnote{https://exofop.ipac.caltech.edu/tess/}, which contains the TIC id of the confirmed and known planets to date, and all the parameters of the planet. In this database, there are roughly 147 systems labeled as Confirmed Planets (CP), which were confirmed as such after the TESS observations were performed; and 420 systems labeled as Known Planets (KP), which were discovered previous to the TESS observations. As second dataset, we use the light curves previously labeled as planet candidate (PC) by DL models, which correspond to the data labeled by \citet{yu2019identifying} and \citet{rao2021nigraha}(Nigraha), that were disclosed in a public repository\footnote{https://github.com/yuliang419/Astronet-Triage/blob/master/astronet/tces.csv}. We filter the light curves with greater confidence with a threshold $>0.7$, of which we discarded 23\% of them since they corresponded to light curves previously labeled as FP. For this second dataset we collected 153 TCEs. \par 


The objective of our model is to detect the differences between the characteristics of planetary transit signals from those of typical false positives, such as eclipsing binaries (EBs), background EBs, stellar variability (V), or Instrumental systematic (IS). In that sense, for the case of false positive signal samples, we include information available in the ExoFOP-TESS catalog, where we collected a total of 224 EBs and 6 variable stars, and 7 candidates labeled as instrument systematics. We also included 1,886 light curves from the \citet{yu2019identifying} work, which correspond to 711 EBs, 520 IS, and 655 V.\par

Additionally, we include to our dataset some non-transit samples that were not previously classified as TOI's, and not belonging to the Nigraha or \citet{yu2019identifying} PCs. For this, we use a method for detecting time-series planetary transits, the Transit Least Squares (TLS) algorithm proposed by \citet{hippke2019optimized}. This method has already been successfully applied by \citet{rao2021nigraha} to identify candidates and compute the transit parameters. In addition to the transit parameters, TLS tool\footnote{https://transitleastsquares.readthedocs.io/\_/downloads/en/latest/pdf/} returns the signal to noise ratio (SNR), and the signal detection efficiency (SDE) related to the value of the TLS power spectrum. SDE has become a standard metric with successful results in the exoplanet hunting community \citep{hippke2019optimized}. To find signals that do not correspond to planetary systems, firstly, we compute the transit parameters on random samples from sectors 1-6. Secondly, we compute the peak in the TLS power spectrum to select the light curves with a threshold of $SNR < 6$. Finally, we filter to include those that have already been previously classified by \citet{yu2019identifying} as J (Junk). This class includes a mix of V and IS, of which we collected a total of 3,000 samples. This way, we obtain samples with greater confidence of belonging to non-transit signals. \par

The final dataset contains 720 true transit signals and 5,123 signals associated with false positive scenarios. We then split our dataset into three subsets: i) 80\% for the training set, ii) 10\% for the validation set and iii) 10\% for the testing set, which is dedicated to measure the performance of our model of which 15.4\% are positive instances. As we can see, our dataset is highly imbalanced, so we work with specific metrics to evaluate the performance of our model (see Section ~\ref{performancemetrics}).

\subsection{Data augmentation}
\label{sec:data_augmentation}


Data augmentation is a technique used to augment data from the same data set. This process consists of modifying random samples from the training set. We take this technique to augment our training data. For this process we consider two techniques. First, we vary the $t_{min}$ and $t_{max}$ of the transit view following the same procedure described in Section~\ref{sec:preprocessing_flux}. We perform this process on a subset of training data corresponding to 50\% of the samples. As second step, similar to \citet{ansdell2018scientific} and \citet{osborn2020rapid}, we add white noise with a standard deviation that is randomly chosen from a uniform distribution between 0 and the mean of the flux to random samples during training. We follow the same steps for each time series of the centroids. This means that the model is trained with different samples in each epoch, preventing overfitting and improving generalization. Adding white noise to signals has been shown to achieve signal variability without changing transit patterns and a better performance of the model (\citet{ansdell2018scientific}; \citet{pearson2018}; \citet{zucker2018shallow}; \citet{osborn2020rapid}). Regarding the stellar and transit parameters, these do not have any change during the data augmentation process since the techniques we adopt do not affect the characteristics of the star, such as $T_{eff}$, [Fe/H], depth or duration of the transit. \par

\section{Training Process and Performance Evaluation}
\label{sec:training}

In this section, we describe the implementation, and training process, including the regularization and optimization techniques we adopted to implement our model. Finally, we describe the metrics we adopted to evaluate the performance of our model.

\subsection{Implementation}
Our model described in this work has been implemented in Pytorch (\citet{paszke2019pytorch}) an open source machine learning framework. All experiments are conducted using NVidia GeForce 1080 Ti with 11GB of memory. The code is available at \hyperlink{https://github.com/helemysm/attention\_exoplanet}{$\texttt{https://github.com/helemysm/attention\_exoplanets}$}

\subsection{Training}

We train our model using the labeled datasets described in Section~\ref{datasets}. Our model has exoplanet transit and false positive signals labels $(1,0)$ as output options. 

We perform the data augmentation process described in Section~\ref{sec:data_augmentation} for the training set only, i.e. to 80\% of all samples. This training set does not have augmented data from the testing set, and furthermore, the samples for model evaluation do not have any augmentation process. Finally, to train our model, we use all samples in batches of 100 over 60 epochs to learn features of a transit signal.

\subsection{Regularization \& Optimizers}

Dropout is a regularization technique, where some neurons are randomly deactivated during the training phase \citep{srivastava2014dropout}. We used this technique the same way as in \citet{vaswani2017attention}. We apply a dropout of 0.1 on the output of each layer, to the positional encoding of the local and global view encoders.

The ADAM algorithm is a variant of the stochastic gradient descendent (SGD) \citep{kingma2014adam}. Using Adam instead of SGD allows the learning rate to change as the network learns. To initialize each model, we set a high learning rate of $\alpha =$ 0.001, which decreases as the network learns and the value of the loss function decreases. The idea of starting the training with a high learning rate allows the model to start learning faster, but with the risk of skipping the optimal solution \citep{goodfellow2016deep}. For this reason, we reduce the learning rate as the model learns, allowing it to learn a more optimal set of weights. To achieve this, we introduce a learning schedule, which is beneficial for reducing the learning rate \citep{goodfellow2016deep}. The initial learning rate of $\alpha =$ 0.001, decreases by 20\% if after 5 or 10 epochs the performance has not been improved. \par

\subsection{Performance Evaluation}
\label{performancemetrics}
In machine learning, classification is the task of predicting the class to which the input data belongs. In this sense, to evaluate how well the classifier manages to predict, evaluation metrics are typically used to reflect the true performance of the classifier. \par

We evaluate our model using two classes: planet (positive instances) and not planet (negative instances which correspond to false positive signals). As we mentioned before, most of our dataset belongs to false positive signal. Therefore, we evaluate the performance of our model with evaluation metrics for unbalanced datasets. Confusion matrix, recall, precision and F1-score are metrics that consider an unbalanced dataset. Thus, the higher the recall and precision value, the better the performance of our model \citep{ivezic2014statistics}. Below we describe the metrics for measuring performance based on the number of TP (true positive value, number of positive samples classified correctly), FN (false negative value, number of actual positive samples classified as negative), and FP (false positive value, number of actual negative samples classified as positive instance), which we use to evaluate our model:


\begin{itemize}

\item \textbf{Confusion Matrix}: displays a matrix with the performance summary of a classification model \citep{stehman1997selecting}. This summary shows the percentage of correct and incorrect predictions for each class. Each row of the matrix represents the prediction of the classes (x-axis), while the columns represent the true label (y-axis). Thus, we obtain information about the prediction errors of our model with respect to each class.  

\item \textbf{Recall}: from all the true transit signals that exist in our testing dataset, how many are predicted as transit signal. As our classification is binary, the recall is defined as:
\begin{equation}
\mathrm{Recall = \frac{TP}{TP+FN} }
\end{equation}
In most cases of binary classification with unbalanced datasets, the recall is the best metric to avoid overfitting.

\item \textbf{Precision}: from the cases that our model predicts true transit signal, how many actually are transit signals. In the same way, from the cases that our model predicts no transit signal, how many actually are no transit signal.
\begin{equation}
\mathrm{Precision = \frac{TP}{TP+FP} }
\end{equation}
In general, the precision is a quantitative measure of the samples that are correctly classified.

\item \textbf{F1-score}: corresponds to the weighted average of recall and precision:
\begin{equation}
\mathrm{F1_{score} = 2 \times \frac{Precision \times Recall}{Precision + Recall} }
\end{equation}
Considers recall and precision, where a high value indicates good performance in binary classification.

\item \textbf{Precision-Recall curve}:
One of the most important metrics to evaluate the performance of an unbalanced classification model is the precision-recall curve, which summarizes the trade-off between precision and recall for different thresholds.

\end{itemize}

\begin{table}
	\centering
	\vspace{5mm}
	\caption{Summary of the results in terms of precision, recall and F1-score of our best model on the test set. The standard deviation $\sigma$ for all three metrics is less than 0.005.}
	\label{tab:result_performance_table}
	\vspace{4mm}
	\begin{tabular}{lccccr} 
		\hline
		\textbf{Class} & & & \textbf{Precision} & \textbf{Recall} & \textbf{F1-score} \\
		\hline
		\hline
		not planet & & & 0.96 & 0.96 & 0.96\\
		planet & & & 0.80 & 0.80 & 0.80 \\
		\hline
		\vspace{1mm}
	\end{tabular}
\end{table}

\section{Results}
\label{sec:results}

Our model was evaluated with samples, which the network has not seen up to this point. We evaluate our model with metrics described in Section ~\ref{performancemetrics}. As mentioned before, our dataset is highly imbalanced. In view of this, we focused on evaluating our model based on the recall precision and the F1-score.

\begin{figure}
\begin{center}
\vspace{5mm}
\includegraphics[width=0.38\textwidth]{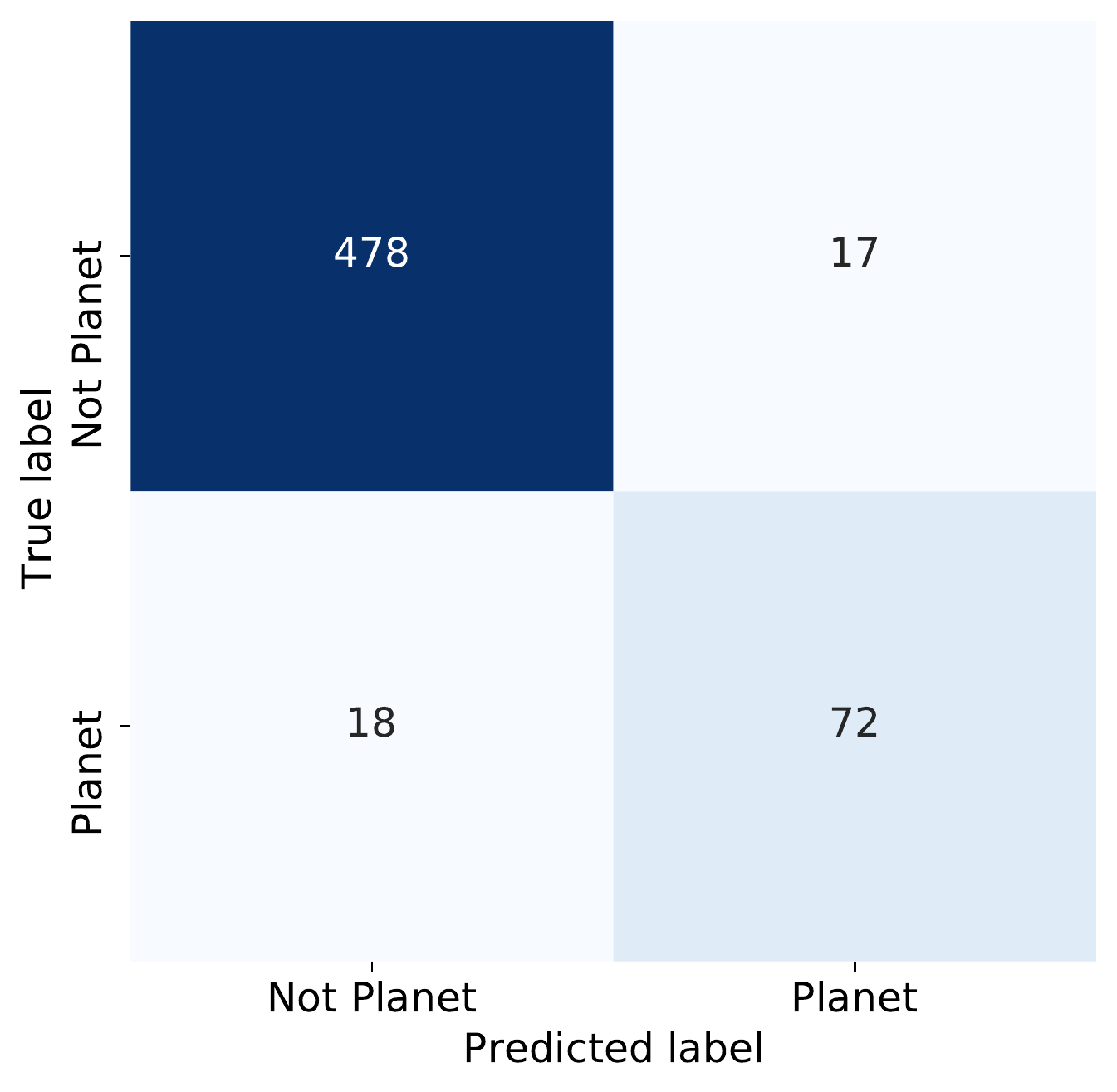}
\vspace{3mm} 
\caption{\label{fig:cm_real}
Confusion matrix resulting from the application of our architecture on the evaluation dataset that contains confirmed planets CP and known planets KP described in Section~\ref{datasets}. Planet corresponds to positive instances and no planet to negative instances. }
\vspace{0mm}
\end{center}
\end{figure}

For each evaluation in all of the experiments, we perform 10-fold stratified cross-validation taking the mean as the final result. Table~\ref{tab:result_performance_table} shows the result of the evaluation. The table shows that our model correctly classified 88.0\% of the samples in total. We computed confusion matrices to display the percentage of correct and incorrect predictions. Figure~\ref{fig:cm_real} shows the confusion matrix for the test set, where 80\% of the instances with planet labels were correctly classified. \par


We compared our model with \citet{rao2021nigraha}. We recreated the Nigraha architecture with the parameters described in \citet{rao2021nigraha}. Then, we train and evaluate it with the same dataset described in this work. As a result, Nigraha correctly classified 86.6\% of the samples in total, and with a recall of 77.7\% for the samples with planet labels. In this way, we compared two strategies that use different DL models. Additionally, we compared our architecture against traditional models, such as support vector machine (SVM) and Multilayer perceptron (MLP). In all the experiments performed, both models do not exceed a precision with a threshold $>0.6$. Therefore, like \citet{Shallue_2018} and \citet{pearson2018}, architectures based on convolutional layers and, in our case, self-attention mechanisms achieve superior results for this type of time series.\par

Nonetheless, beyond getting better performance, we focused on detecting the characteristics of an exoplanet transit and of a false positive signal. In the ablation study (see Section ~\ref{ablation_study}), we show the contribution of each component of our model, and we also observe how the centroid information improves the performance.  \par

\begin{figure}
\begin{center}
\vspace{4mm}
\includegraphics[width=0.49\textwidth]{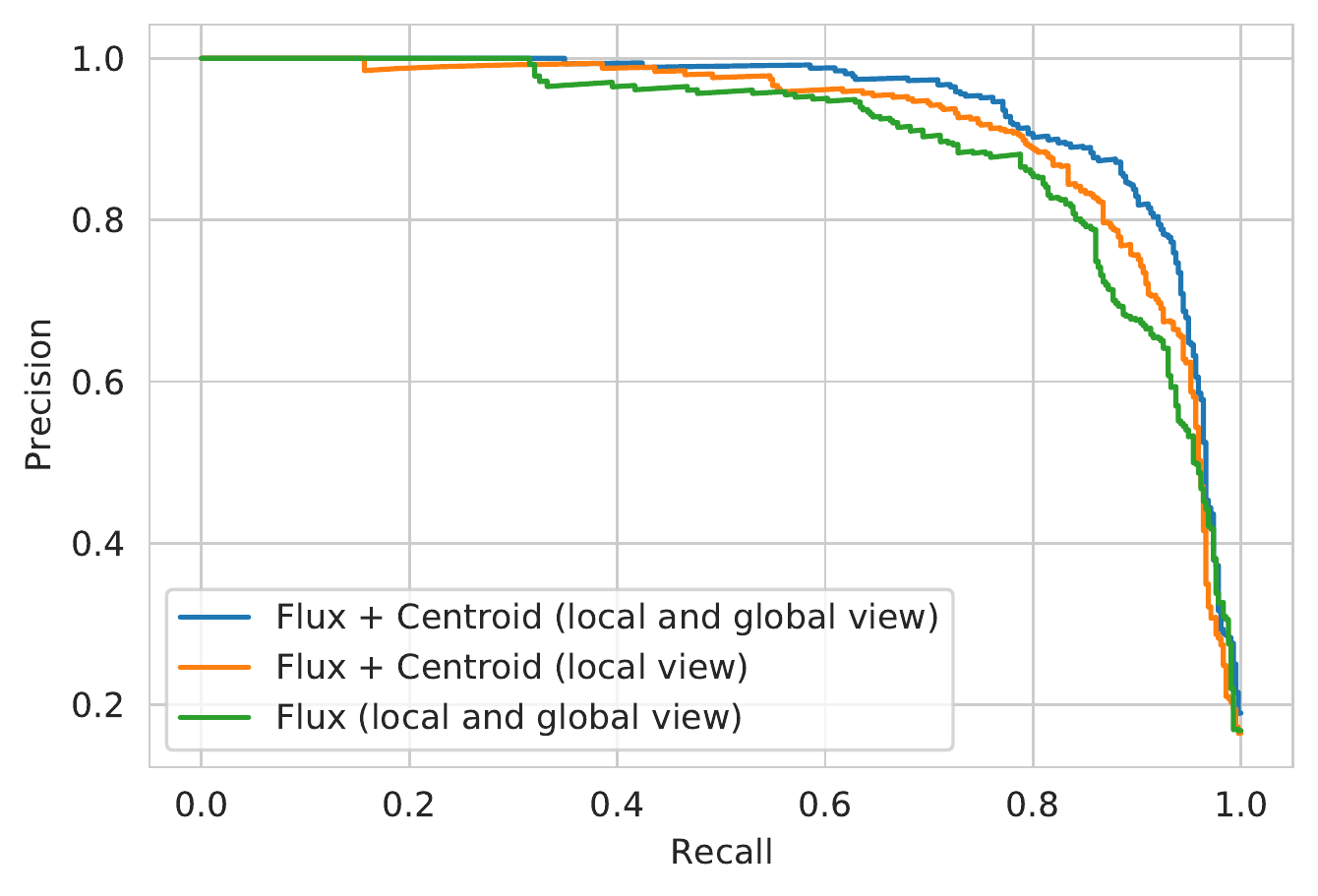}
\vspace{1mm} 
\caption{\label{fig:prec_recall}
The precision-recall curve for test sets. The figure shows better performance if we include the centroid information in the model training. Furthermore, we can see a slight difference between training the model with only the local view of the transit + centroid compared to training the model without the centroid but with both views.}
\vspace{-1mm}
\end{center}
\end{figure}

\begin{table}
	\centering
	\vspace{5mm}
	\caption{Ablation study of different components of our model. Default model is: $PE = {time}$, $d_{emb} = 512$, $h = 4$, $l = 4$ with local and global view and centroid information. }
	\label{tab:ablation_table}
	\vspace{4mm}
	\begin{tabular}{lccccc} 
		\hline
		\textbf{Description} & \textbf{Avg. Precision} & \textbf{Avg. Recall} & \textbf{Avg. F1-score} \\
		\hline
		\hline
		Default & 0.880 & 0.880 & 0.880 \\ %
		\hline
		w/o $PE$   &0.495 & 0.430 &  0.425  \\ 
		$PE_{abs\_time}$  & 0.875 & 0.870 & 0.870 \\ 
		\hline
		$l = $ 1     & 0.860 & 0.865 & 0.860 \\ 
		$l = $ 8     & 0.885 & 0.880 & 0.880 \\ 
		\hline
		Without centroid    &  0.850 & 0.860 & 0.850 \\ 
		Without global view   &  0.865 & 0.870 & 0.870 \\ 
		\hline
		$d_{emb} = 128$   &  0.845 & 0.855 & 0.850   \\
		$d_{emb} = 256$     & 0.865 & 0.870 & 0.870  \\ 
		\hline
	\end{tabular}
	\vspace{3mm}
\end{table}

\subsection{Ablation study}
\label{ablation_study}
To understand the contribution of the most important components of our architecture, we evaluate the performance with ablation experiments. It is important to mention that all our experiments consider the information on the stellar/transit parameters since the objective is to observe the relative importance of each parameter. Table~\ref{tab:ablation_table} shows the performance of our model based on the number of heads and layers. In addition, we can observe the impact of incorporating the centroid into the encoder input with the flux time series. Specifically, we train our model by ablating specific components:

\begin{itemize}

\item \textbf{Importance of positional information}: 
We used three types of positional encoding (PE): no positional encoding, a sinusoidal encoding that depends on observation time, and sinusoidal encoding that relies on the absolute position. Since the attention is calculated at each time step independently, the sequential order of the observations is lost. Without the sequential order, the signal of a transit would have observations with the value of the brightness decrease in any time step of the sequence, making it challenging to identify the transit patterns in relation to time. We evaluated the performance if we did not inject a sequential order into the input, and the average precision performance of the model was 50\%, similar to a random result for binary classification. Furthermore, we note that the type of positional encoding slightly influences the results. A PE that depends on the observation time performs $\sim$1.0\% better than a PE that considers the same space between tokens. With these results, we found that both encoding produce very similar results due to the fact that the observation time variation is minimal without achieving significant changes in the positional encoding vector. \par

\item \textbf{Number of layers and attention heads}: 
In all our experiments we use the same number of layers and heads for each encoder. To choose the appropriate number of heads, we study the entropy of each head in each layer (see Section ~\ref{individual_attention_heads}), where we considered 8 heads. We noticed that there were no significant changes in the model's performance, either with $h = $8  or $h = $4. However, the first 4 heads have the lowest entropy, meaning that each head's attention is less dispersed and more focused on specific observations of the sequence. As we mentioned before, in multihead attention theory, the heads are specialized in different parts of a sequence, but also, some layers can be reduced to only one head without changing the performance of the model \citep{michel2019sixteen}.  \par

Our default model is trained with 4 heads and a varying number of layers, since there were no significant differences in the results with 8 heads. By reducing $l =$ 1, the model performance evaluation metrics decreases 1.5\%. In the case of working with $l = $ 8, the model has a similar performance to $l = $ 4, where basically there is a greater difference in the model training time. 

\item \textbf{Without centroid Information}: We perform experiments without considering the centroids. As a result, the model reduces its performance by 2\%. So, we achieve better results if we include this information in the encoder. In addition, the attention weights are more clearly focused when the planetary transit begins and ends, which benefits the interpretability of the model decision (see Section ~\ref{attention_map_analysis}).

\item \textbf{Without global view}: We perform experiments without considering the global view of the transit signal. In the experiments, we observed that without including the global view of the transit signal, the performance of the model decreases slightly by 1.0\%, but not significantly.

\vspace{5mm}

Figure~\ref{fig:prec_recall} shows the precision versus recall for three different models: i) considering the information of the centroids and both views (local and global), ii) local view and centroids, iii) local and global view, but without centroids. In the figure, we observe that the architecture that considers the information of the centroids and both views (local and global) performs better. In the case of the architecture that considers the centroids but not the global view, it behaves with an improvement compared to the architecture without centroids but with both views.

\end{itemize}

\vspace{2mm}

\subsection{Individual Attention Heads}
\label{individual_attention_heads}

To analyse each attention head, we use the attention entropy described in Section~\ref{sec:attention_entropy_model}. The entropy is measured in $\mathit{nats}$ units, which means ``The natural unit of information'', and we use it to measure the concentration of the attention distribution at a time $t$. An entropy above the average indicates that the attentions are more dispersed. Otherwise, and if the entropy values of each head vary, it could indicate that the attention of those heads is focused on specific observations. \par

\begin{figure}
\begin{center}
\vspace{5mm}
\includegraphics[width=0.44\textwidth]{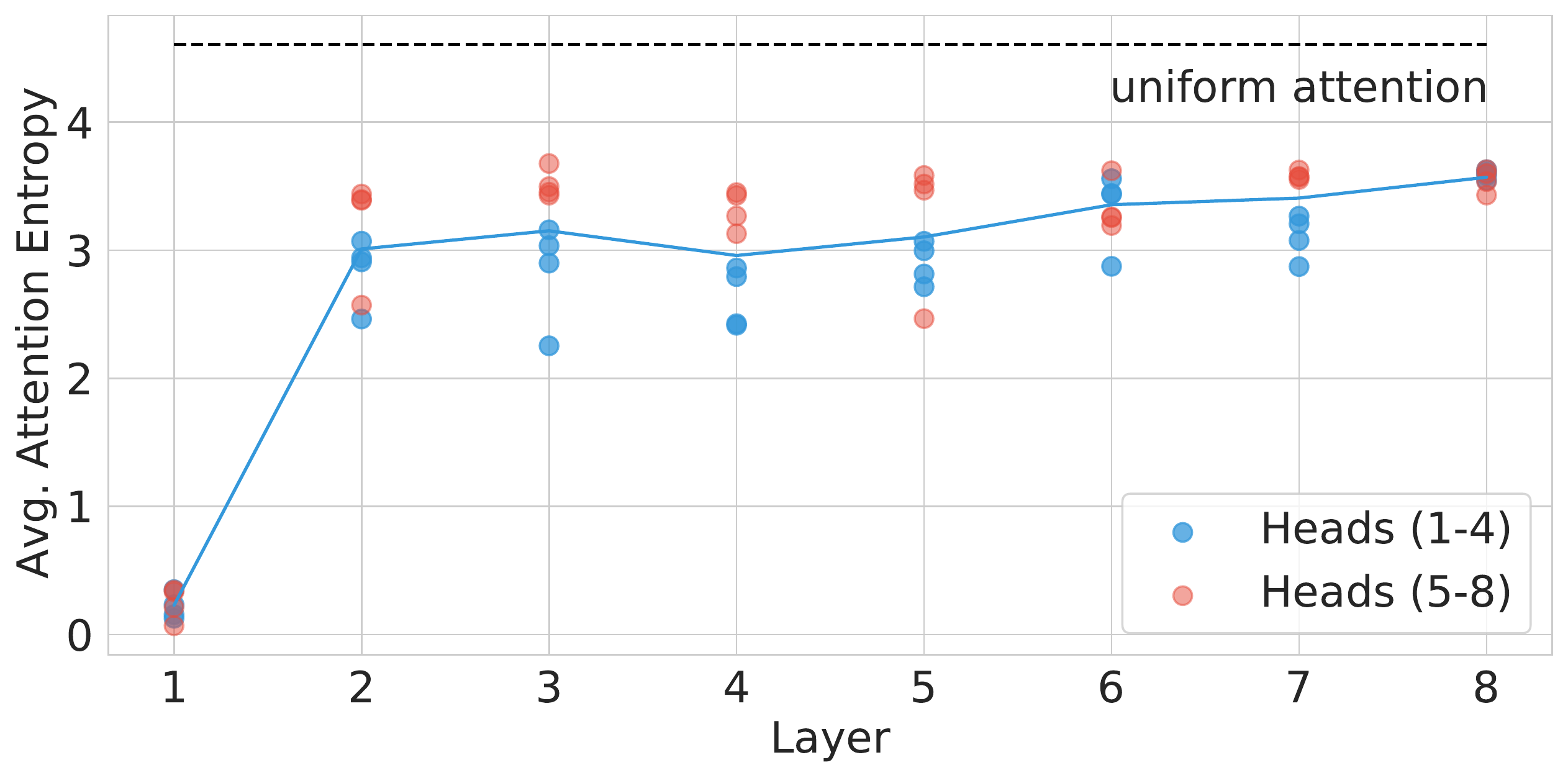}
\vspace{2mm} 
\caption{\label{fig:entropy}
Average entropy of attention distributions. The blue line indicates the average attention entropy calculated for each layer, where each layer has 8 attention heads. The points represent the average value of the attention entropy of each head. The first four heads (1-4) of each layer are the blue dots, and the red dots correspond to the last four heads (5-8).}

\vspace{2mm}
\end{center}
\end{figure}

Figure~\ref{fig:entropy} shows the average entropy, where we observe that some attention heads, especially in the last layers (6, 7 and 8), have wide attention, but also the heads have similar entropy values. This means that in those layers, there is more attention. These attention heads generally have at most 10\% of their attention weights on a single token (observation). Furthermore, in Figure~\ref{fig:entropy} we show the attention entropy of the heads in each layer, where the first heads (1-4 blue dots) have below average entropy compared to the last heads (5-8 red dots). This means that the attention weights are concentrated on specific observations. This analysis helps us define the optimal number of heads for our model, in our case, $h = $ 4. Whit $h =$8 there were no significant differences in model performance. Another option for head analysis is to perform a process called ``head pruning'' on each layer \citep{voita2019analyzing}. This means that heads can be removed and will not significantly affect the result. In future work, we will consider a combination of techniques, including head pruning, in such a way that we can improve the analysis of the model results.


\subsection{Analysis of Attention Maps}
\label{attention_map_analysis}

\begin{figure*}
\begin{center}
\includegraphics[width=1\textwidth]{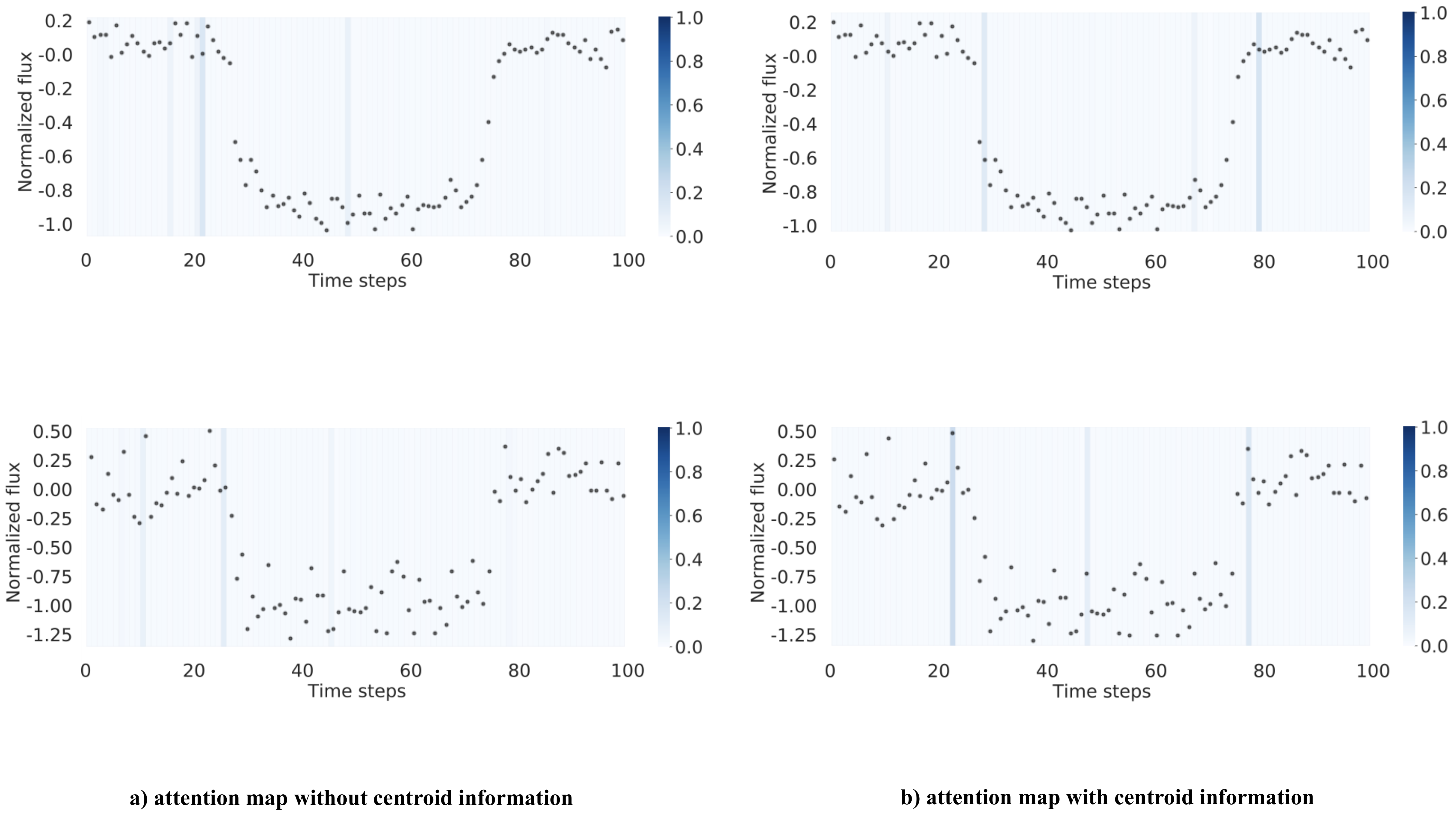}

\vspace{5mm} 
\caption{\label{fig:att_map_examples}
Interpretation of predictions for two different planetary transit signals from the test set. Each plot represents the transit signal (gray dots), with the attention weights that the model calculated and the bar on the left represents the proportion of the associated weight (Heatmap of the self-attention weight matrix). On the left in a), the two transits are shown with the attention map for the trained model without the information of the centroids. On the right in b), the two transits are shown with the attention map for the model trained with the information of the centroids. Notice that in a), the weights of attention are more dispersed, whereas in b) the weight of attention is stronger when the transit begins and ends. Uniform weights are observed for the other timesteps.}

\vspace{1mm}
\end{center}
\end{figure*}

We interpret the importance of the signal characteristics for the final prediction by analysing the weights of attention. To do this, we analyse the attention head with an average entropy value over all the layers. Then, we interpret the distribution of attention based on the average attention described in Equation~\ref{attention_weights}. \par

We take two random samples from the test set in order to observe the attention map for the model trained without the information of the centroids, and for the model trained with the concatenated representation of the flux and centroid. Figure~\ref{fig:att_map_examples}(a) on the left shows the two exoplanetary transit signal samples of the trained model without the centroids. We observe that the distribution of attention weights is focused at various transit times. In other words, the attention weights are more dispersed. In the case of the model trained with the information of the centroids concatenated to the flux, the Figure~\ref{fig:att_map_examples}(b) on the right shows that the attention weights are focused when the planetary transit begins and ends. \par

Beyond slightly improving the performance of the model, including the centroid information makes the attention weights focus when the brightness decreases and increases, i.e. we obtain a better input embedding for the model. This means that in many cases, the analysis of the centroid information could be essential to identify a planetary transit signal. In others words, the probability that a signal belongs to a planetary transit signal increases if it does not have false positive characteristics. This interpretation is similar to the analysis that experts perform to confirm a candidate.\par

\subsubsection{Transit signals of planets}
\label{attention_map_tpositives}

\begin{figure*}
\begin{center}
\vspace{6mm}
\includegraphics[width=1\textwidth]{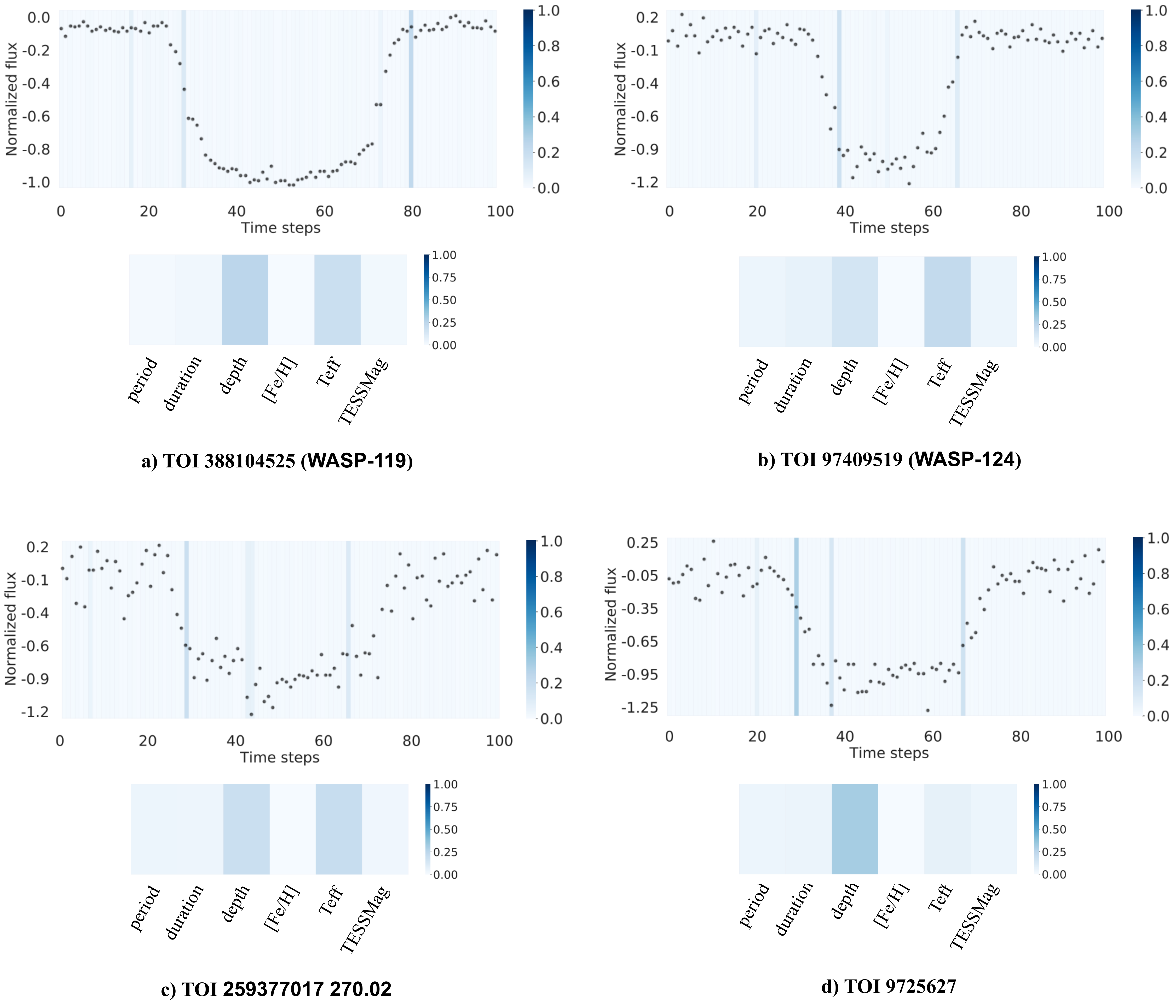}

\vspace{5mm} 
\caption{\label{fig:att_map_transit_params_true}
Interpretation of predictions for four signals of a planetary transit of TOI's labelled as KP or CP. Each sub-figure shows the transit signal (gray dots) with the attention weights that the model calculated in the figure above and the stellar/transit parameters in the heat map below.}
\vspace{1mm}
\end{center}
\end{figure*}

Figure~\ref{fig:att_map_transit_params_true} shows four true positive samples, i.e. the signal of a planetary transit of TOI's labelled as KP or CP and their stellar/transit parameters. Our first observation is that there is more attention when the transit starts, and this attention continues until the end of the sequence of observations. In addition to the beginning of the transit, there is also attention when the transit ends. Another characteristic to highlight are the attention patterns towards the observations before and after the transit, which have very low attention and are sometimes ignored. This is because these observations are not relevant to the final decision of the model. This analysis is very similar to the interpretation of a text in NLP, since there are irrelevant words that do not affect the meaning of a paragraph. Regarding the stellar/transit parameters, the four samples in the figure show the attention map for the six parameters that we consider. In these figures, we observe that the decision of whether the signal belongs to a planetary transit is based mostly on the values of the transit depth and the temperature of the host star. In contrast, the attention is low for the [Fe/H] parameter in all examples. \par

\subsubsection{Transit signals of false positives}
\label{attention_map_tpnegatives}

\begin{figure*}
\begin{center}
\vspace{5mm}
\includegraphics[width=1\textwidth]{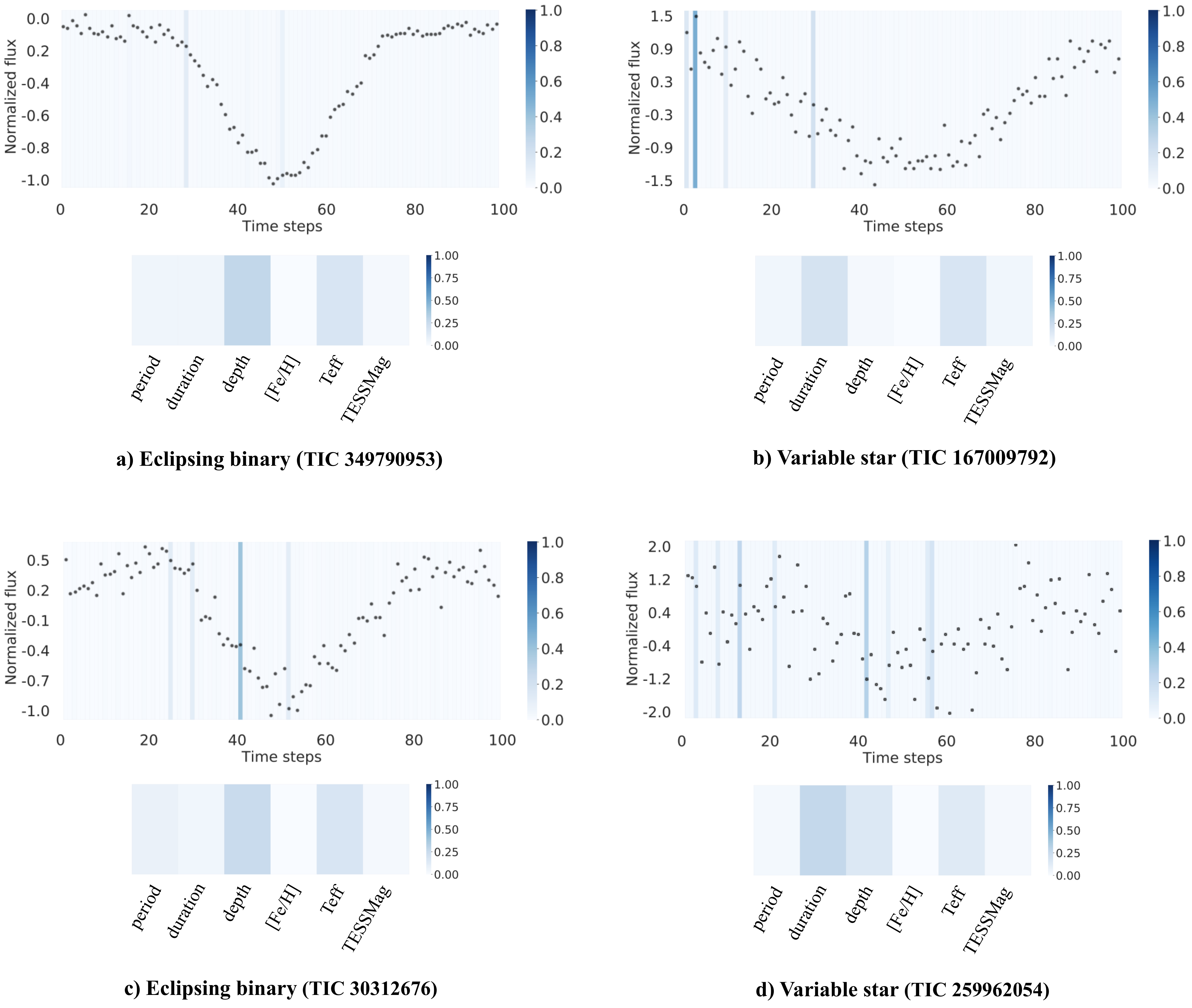}
\vspace{5mm} 
\caption{\label{fig:att_map_transit_params_falsep}
Interpretation of predictions for four signals, which was correctly classified as not planet, which correspond to TOI's labelled as false positives. Each sub-figure shows the signal (gray dots) with its background attention heat map (blue), and the attention map of its stellar/transit parameters below.}
\vspace{1mm}
\end{center}
\end{figure*}

Figure~\ref{fig:att_map_transit_params_falsep} shows four samples of false positive signals. Specifically, two of these samples belong to an EB; (a) and (c), and two belong to variable stars; (b) and (d). In the case of EB samples, we observe that the attention is focused close to the beginning and at the minimum point of depth of the transit. Also, in the attention map of its stellar/transit parameters, the depth value shows more attention than average. The distribution of these attentions is similar to the characteristics of EBs since their transits usually have greater depth than a planetary transit. It is the relevant characteristic that the model detected to distinguish this type of false positive. In the case of, Figure~\ref{fig:att_map_transit_params_falsep}(b) belongs to another sample of a false positive signal, a variable star sample. In this sub-figure, we observe that the attention weights are more significant at the beginning of the light curve, which could seem like the beginning of some planetary transit but with a longer duration. However, that attention ends before the maximum point of depth. This is related to the attention to the stellar/transit parameters since the attention is concentrated on the duration of the transit. Finally, the last sample (d) belongs to another variable star, this sub-figure shows a more dispersed distribution of attention weights and this dispersion is concentrated until the maximum point of transit depth. This means that the model has found the most important features up to that point and thus classifies it as a non-planet. \par

\vspace{3mm} 

In general, the attention maps help us to indentify the patterns that the model detected to characterize planetary and false positive signals. All the attention of the model is distributed in each datapoint, and most of the datapoints have low attention. This is because the model determines they are irrelevant to characterize a transit signal. The datapoints with the most attention are those that represent the decrease or increase in brightness of the star in a significant way. \par

\begin{figure*}
\begin{center}
\vspace{5mm}
\includegraphics[width=0.9\textwidth]{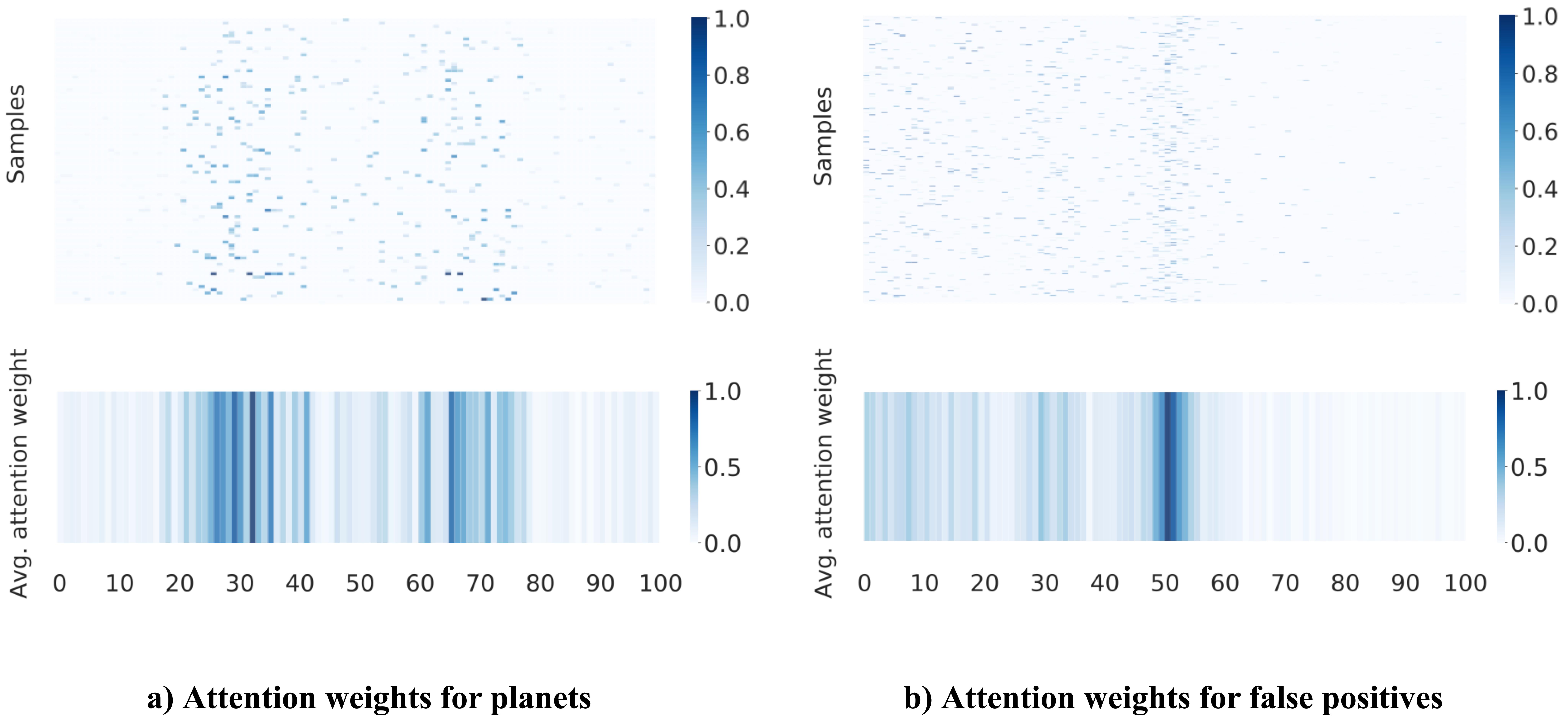}
\vspace{5mm} 
\caption{\label{fig:attention_trend}
Attention maps of transit signal samples. The graphs overhead are the attention maps of all the test samples set. The x-axis in a heat map representing time step. The charts down show the general attention trend for each time step, with warmer colors indicating a higher concentration of weights at specific times.}
\vspace{1mm}
\end{center}
\end{figure*}

 Figure~\ref{fig:attention_trend} shows the attention maps of all the test samples set for planetary and false positive signals, and below the heat maps that show the general trend of all the samples for each time step. Figure~\ref{fig:attention_trend}(a) clearly shows the patterns that characterize planetary transit signals, where the transit begins and ends. Then, Figure~\ref{fig:attention_trend}(b) shows slightly more dispersed attention. However, the pattern of attention is concentrated close to the time in which there is a greater depth of transit. \par

Analysis of the attention maps showed that it is possible to distinguish observations containing the main features of a planetary transit and a false positive signal. Mainly, we remember that the depth of the transit is an essential characteristic that exoplanet hunters typically analyze to examine false positive signals. In this sense, our results regarding the analysis of the attention maps showed that the model also observes this depth characteristic to make a decision. With this in mind, working a light curve with a sequential model instead of a CNN has advantages for a better understanding of the predictions. Moreover, it is possible to analyze other parameters in addition to the light curve, such as stellar or transit parameters. This analysis would help us better understand the correlation of the stellar/transit parameters with the prediction and, thus, facilitate the candidate examination process. \par

\section{Conclusions}
\label{sec:conclusions}


In this work, we present a DL architecture for classifying the light curves of transiting exoplanet candidates. Furthermore, through our architecture, it is possible to interpret a prediction to distinguish signals of planetary and false positives. The interpretation of the prediction consists of showing the parameters and observations that the model pays attention to classify a transit-like signal. In addition to the interpretation, the implemented architecture is able to achieve comparable results to state-of-the-art architectures for the identification of exoplanetary signals. \par

While this work is motivated by the problem of the classification of signals of planetary transits, the architecture is adequate to interpret and observe the importance of each element of the input. Previous works have shown that model performance improves with the inclusion of centroids. In our work, with the help of the attention interpretation, we also show how it affects the model to improve performance, which we think is also important. In addition, to obtain a better interpretation of the model's attention, we analysed the attention entropy of the heads. This helps us determine the number of heads for implementing a model based on attention mechanisms.

The attention mechanism is a theory that has been proven to be successful in NLP applications but has not been previously explored in exoplanet science, where much of the analysis is from sequential data such as light curves. Models like the Transformers are designed to work in complex scenarios for CNNs and RNNs, like time series, since the greater the number of observations of the light curve, the greater the complexity of the model. In this sense, working with architectures such as the Transformer is more beneficial if we consider light curves with more observations, overcoming the limitations of other models. For example, the light curves from telescopes designed to detect transiting planets are characterized by long light curves. We hope that applying this approach can improve other astronomical event classification tasks. \par

Our future work aims to scale up to longer light curves without building phase-folded light curves, which are typically used to train a DL model. For example, to classify entire light curves and predict whether or not they belong to a star that hosts exoplanets since a light curve could contain one or more exoplanet candidates. Furthermore, we can take advantage of the interpretability of attention mechanisms since analysing the distribution of attention weights could indicate where the possible transit signals are located.\par


\section*{Acknowledgements}
The authors would like to acknowledge the support from the National Agency for Research and Development (ANID), through the FONDECYT Regular project number 1180054. HS acknowledges support from ANID, through Scholarship Program/Doctorado Nacional/2020-21212185. FP acknowledges support from ANID, through Scholarship Program/Doctorado Nacional/2017-21171036. DM acknowledges the support from ANID by Proyecto Basal FB210017, through the National Center for Artificial Intelligence CENIA. This paper utilizes public information of the TESS and Kepler mission. Funding for the TESS mission is provided by the NASA Explorer Program. Finally, thanks to the availability of data from TESS, we were able to examine and display our analysis of transit signals.

\section*{DATA AVAILABILITY}

All the data referenced here and that was used is public information about the TESS mission from Mikulski Archive for Space Telescopes (MAST) \footnote{https://archive.stsci.edu/tess/bulk\_downloads/bulk\_downloads\_ffi-tp-lc-dv.html}. For the original light curve dataset, the reader is advised to explore the available documentation on the TESS dataset \citep{tenenbaum2018tess}.

\bibliographystyle{mnras}
\interlinepenalty=10000






\bsp	
\label{lastpage}
\end{document}